\documentclass[fleqn,usenatbib]{mnras}

\usepackage{newtxtext,newtxmath}
\usepackage[T1]{fontenc}

\DeclareRobustCommand{\VAN}[3]{#2}
\let\VANthebibliography\thebibliography
\def\thebibliography{\DeclareRobustCommand{\VAN}[3]{##3}\VANthebibliography}

\usepackage{graphicx}	
\usepackage{amsmath}	

\title[Hemispheric analysis of active regions]{Hemispheric analysis of the magnetic flux in regular and irregular solar active regions}

\author[A. Zhukova]{
A. Zhukova,\thanks{E-mail: anastasiya.v.zhukova@gmail.com}
\\
Crimean Astrophysical Observatory, Russian Academy of Science, Nauchny, Bakhchisaray 298409, Crimea
}

\date{Accepted XXX. Received YYY; in original form ZZZ}

\pubyear{2024}

\begin{document}
\label{firstpage}
\pagerange{\pageref{firstpage}--\pageref{lastpage}}
\maketitle

\begin{abstract}
Studying the hemispheric distribution of active regions (ARs) with different magnetic morphology may clarify the features of the dynamo process that is hidden under the photospheric level.
The magnetic flux data for 3047 ARs from the CrAO catalog between May 1996 and December 2021
(cycles 23 and 24) were used to study ARs cyclic variations and perform correlation analysis.  
According to the magneto-morphological classification (MMC) of ARs proposed earlier,  
subsets of the regular (obeying empirical rules for sunspots) and irregular (violating these rules) ARs were considered separately.
Our analysis shows the following. For ARs of each MMC type, in each of the hemispheres, time profiles demonstrate a multi-peak structure.
The double-peak structure of a cycle is formed by ARs of both MMC types in both hemispheres.
For the irregular ARs, the pronounced peaks occur in the second maxima (close to the polar field reversal). Their significant hemispheric imbalance might be caused by a weakening of the toroidal field in one of the hemispheres due to the interaction between the dipolar and quadrupolar components of the global field, which facilitates the manifestation of the turbulent component of the dynamo. The similarity of the irregular ARs activity that was found in adjacent cycles in different hemispheres also hints at realization of the mix-parity dynamo solution. For the quadrupolar-like component of the flux (compiled in the simple axisymmetric approximation), signs of oscillations with a period of about 15 years are found, and they are pronounced specifically for the irregular groups. This MMC type ARs might also contribute in $\alpha$-quenching.

\end{abstract}




\section{Introduction}

Active regions (ARs) are the most famous manifestation of the solar activity, reflecting its 
cyclical nature, which still harbors unclear aspects. \citep{Schwabe44, Hale19}.
For instance, despite decades of research, the origin of the North-South (N-S) asymmetry 
of the solar activity is not fully understood.
The noticeable hemispheric imbalance is confirmed by numerous observations \citep[see, e.g., reviews by]
[]{Hathaway15, Usoskin17}.
This phenomenon is constantly in the focus of modern research on sunspot number, areas 
\citep{Badalyan17, Chowdhury19, Veronig21, Javaraiah22, Batista23, Carrasco23},
and other indicies of the global activity \citep{Xie18, Vokhmyanin22, Taran22, Zhang23, Zhang24, Mursula23}.
The understanding of the N-S asymmetry mechanisms is essential for 
the prediction models \citep{Hathaway16, Lekshmi19, Labonville19, Javaraiah21, Nandy21, Bhowmik23}.

As it is follows from the pioneer magnetic cycle models \citep{Babcock61, Leighton64, Parker55}, 
the activity of the two hemispheres should not differ.
ARs, the source of which is understood as a toroidal component of the global magnetic field,
should occur approximately equally in both hemispheres.
When the magnetic field lines of the global dipole are stretched along the equator in the convection zone 
due to differential rotation, neither hemisphere has an advantage.
On the other hand, the mean-field dynamo theory requires the destruction of the mirror symmetry 
of turbulence in convection zone \citep{Moffatt78, Krause80}, which is a necessary condition for overcoming 
the limitations of anti-dynamo theorems \citep{Cowling33, Zeldovich56}.
Certain theoretical studies assume that the $\alpha$-effect
(which is responsible for restoring the poloidal component of the global field) has a different distribution 
in the two hemispheres \citep[see, e.g., reviews by][and references therein]{Charbonneau23, Karak23}. 

Observational studies allows for different interpretations of the interaction of the two hemispheres.
Some authors suggest that the interdependence of magnetic field systems originating in the hemispheres 
is weak, and the hemispheres are rather independent of each other \citep{Antonucci90, Temmer06, Inceoglu17}. 
However, the noticeable interaction between the hemispheres was reported by \citet{Obridko20, Bisoi20}.
The relationship between the characteristics of adjacent cycles (such as the lag between the activity 
of the hemispheres) and the possible memory of the cycles are also discussed \citep{Chatterjee06,
Zolotova09, McIntosh13, Karak17, Das22}.

The hemispheric distribution of ARs with different  individual properties (observed or simulated) also show peculiarities. For instance, \cite{Mandal16} found that the time profiles of the asymmetry index vary for sunspot groups of different sizes. \cite{Nagy17, Nagy19} showed that the long-term behavior of solar activity (including significant hemispheric asymmetry) considerably depends on the presence of large individual ARs with atypical properties (`rogue' regions). By atypical properties they mean violations of the Hale's polarity law and Joy's law \citep{Hale19}. Recall that anti-Hale ARs demonstrate uncharacteristic (for certain cycle and hemisphere) polarity of the leading spot, while non-Joy groups have unusual tilt, i.e. inclination of the magnetic axis relative to the East-West direction. \cite{Wang14} discussed the possibility of a difference in tilts during the Maunder Minimum and in modern cycles. \cite{Bhowmik19} reported that the tilt randomness is the most crucial element (among diverse components) of the Babcock-Leighton mechanism in resulting hemispheric irregularities in the evolution of polar field.  \cite{Hazra17} revealed the role of anti-Hale ARs in the weakening of the polar field in certain hemisphere, although they found the effect of a single sunspot pair as not very dramatic. \cite{Mordvinov22} shown that the decay of anti-Hale and non-Joy ARs results in the remnant flux surges that are directed towards the pole and transform the conventional order in magnetic flux transport in corresponding hemisphere. 

Distinguishing ARs with atypical properties also underlies the recent magneto-morphological classification (MMC) of ARs \citep{Abramenko18, Abramenko21}. The idea is that, along with a set of the regular ARs (obeying empirical rules for sunspot groups), a special set of the irregular ARs (violating one or more rules) can be considered (see Section \ref{sec:data} for more details). Although both the regular and irregular ARs follow the cycle and supposed to be generated by the global dynamo, the strongest fluxes of the irregular ARs are observed in the second maximum, which may indicate intervention of the turbulent component of the dynamo \citep{Abramenko23}.   As one of the reasons for the irregular ARs hemispheric imbalance, the interplay between the dipolar and quadrupolar components of the global magnetic field was assumed \citep{Zhukova23}. Please note that the role of the quadrupolar component in the solar activity was widely discussed \citep[see, e.g.,][]{Usoskin09, Kapyla16, Shukuya17, Beer18, Karak18ApJ, Schuessler18, Nagy19, Nepomnyashchikh19, Kitchatinov21}. For the large-scale magnetic field, the quadrupole mode was assumed as one of the reasons 
for the N-S asymmetry \citep{Oliver94, Zolotova06, Badalyan11, Zharkova12,Das22}. Grand minima of the solar activity may also be associated with violations of the dipolar parity \citep{Sokoloff94, Olemskoy13, Nagy17, Hazra19}.

In addition, focusing on the totality of ARs-`violators', one may be interested in whether such groups are the result of random deviations that fluctuate the cycle provided by the regular groups, or whether they are an inherent part of the dynamo process that have a special functionality. Although the terminology \citep[for instance, `rogue bipolar magnetic regions' or `irregular ARs', see ][]{Nagy17, Abramenko18, Nagy19, Abramenko21} hints at the former, the peculiar hemispheric asymmetry \citep[found for the irregular ARs number,][]{Zhukova23, Zhukova22MNRAS} encourage us to study this issue in more detail.

Here we analyze the magnetic fluxes of ARs to identify the features of the hemispheric distribution of groups with different magnetic morphology and to reveal their involvement in the dynamo process. Possible signs of the interplay between the dipolar and quadrupolar components of the magnetic field, which might be expressed in the irregular ARs flux profiles, are also considered. The study encompasses two completed solar cycles (SCs), namely, SCs 23 and 24 (from May 1996 to December 2021).	

\section{Data and method}
\label{sec:data}

In this study, unlike our previous works on the N-S asymmetry of the number of sunspot groups
\citep{Zhukova20GA, Zhukova22MNRAS, Zhukova23}, we based on the data on the magnetic fluxes of ARs.
Magnetic fluxes of ARs are widely associated with 
the subphotospheric toroidal magnetic field produced by the global dynamo.
The magnetic flux data (used for each AR once) can be considered as a `generative' activity index 
and allows us to make assumptions about the features of the dynamo process 
\citep{Abramenko18, Nagovitsyn21}.

As a source of the magnetic flux data we used the catalog of the magneto-morphological classes of ARs
of the Crimean Astrophysical Observatory (MMC ARs CrAO catalog). 
The catalog was created in 2017 \citep{Abramenko18, Zhukova18}, and signeficantly redesigned in 2022
in accordance with the approach outlined in \citet{Abramenko21}.
The catalog is available at the CrAO web site (https://sun.crao.ru/databases/catalog-mmc-ars).

Recall that, in accordance with a technique of independent snapshots of full-disk magnetograms
\citep{Abramenko18}, the MMC ARs catalog includes data on ARs that appeared on the disk every 9th day 
in the range of 60 degrees from the central meridian.
Such selection parameters allow satisfying three conditions: 
i) independence of snapshots \citep[as it follows from][a typical correlation time for daily
sunspot series is about 7 days]{Oliver95} and accounting of each unique AR once;
ii) minimizing outcome of the projection effect (ARs with inversion of the magnetic field near the limb 
were discarded);
iii) three 9-day snapshots cover the Carrington rotation (which facilitates subsequent data processing).

For each of the 3047 ARs (that are recorded in the MMC ARs catalog for the period of the study from May 1996 
to December 2021), the catalog contains the calculated unsigned magnetic flux, identifier (NOAA number), 
coordinates and other specific data.
For the SC 23,  the magnetic fluxes of ARs were calculated from the magnetic field data 
of the Michelson Doppler Imager \citep[MDI:][]{Scherrer95} on board the Solar and Heliospheric Observatory.
For the SC 24, the magnetic fluxes were obtained by means of the Helioseismic and Magnetic Imager 
\citep[HMI:][]{Scherrer12} aboard the Solar Dynamics Observatory.
More specifically, the Space-weather HMI Active Region Patches \citep[SHARP:][]{Bobra14} were used 
for the HMI period.
To take into account the systematic difference between the MDI and HMI instruments \citep{Liu12},
during processing of the different cycles data, we applied to them a correction coefficient 
\citep[following][we divided the MDI fluxes by 1.23]{Abramenko23}.
In addition, only those ARs whose magnetic flux exceeding the threshold of $10^{21} Mx$ were considered.
More details regarding the magnetic flux calculation procedure and the features of the current 
version the MMC ARs catalog can be found in \citet{Abramenko23, Zhukova23}.

When analyzing the data, we distributed ARs between two main MMC types.
According to the MMC \citep{Abramenko18, Abramenko21}, all 
studied ARs, except for unipolar spots, were sorted out between the following categories: 
regular -- bipolar ARs obeying empirical laws (rules) for sunspot groups; 
irregular ARs -- all the rest. 
By classical rules we mean the Hale polarity law (implying a certain polarity of the leading 
sunspot depending on whether the AR belongs to the even/odd cycle and N-/S-hemisphere), the Joy law 
(latitudinal dependence for the tilt of ARs) and the prevalence of the leading sunspot rule \citep{Hale19, 
Grotrian50, vanDriel15}.
It is quite obvious that the regular ARs are compatible with classical magnetic cycle
models \citep{Babcock61, Leighton64, Parker55} and the mean-field dynamo theory \citep{Moffatt78, Krause80}.
Their existence is completely determined by the global dynamo action \citep{Abramenko21, Zhukova22GA, 
Zhukova22MNRAS, Abramenko23, Zhukova23}.
The class of the irregular ARs, on contrary, represents a violation of the clear pattern and may indicate 
the interference of other mechanisms of the magnetic field excitation.

In addition, as we interested in the N-S asymmetry of ARs, we distributed ARs 
between N- and S-hemispheres.
As the result, along with the data for all the studied ARs (total unsigned magnetic flux), 
we were dealing with four sets of ARs, depending on their magnetic morphology and location in different 
hemispheres. We also compiled two additional quantities (conventionally named by us `semi-sum' and `semi-difference' parts of the flux, see Section \ref{sec:evenodd}) from the hemispheric data, and that added two more subsets for us to study. We suppose that these parts of the flux might be related to its dipolar-like and quadrupolar-like components, as it is discussed in Subsection  \ref{sec:dipquadr}.

Please note that the flux of each set was calculated separately. Final time series consist of the cumulative magnetic flux data per rotation.

\section{Temporal variations of the regular and irregular ARs in different hemispheres}
\label{sec:temp}

\begin{figure}

\includegraphics[width=\columnwidth]{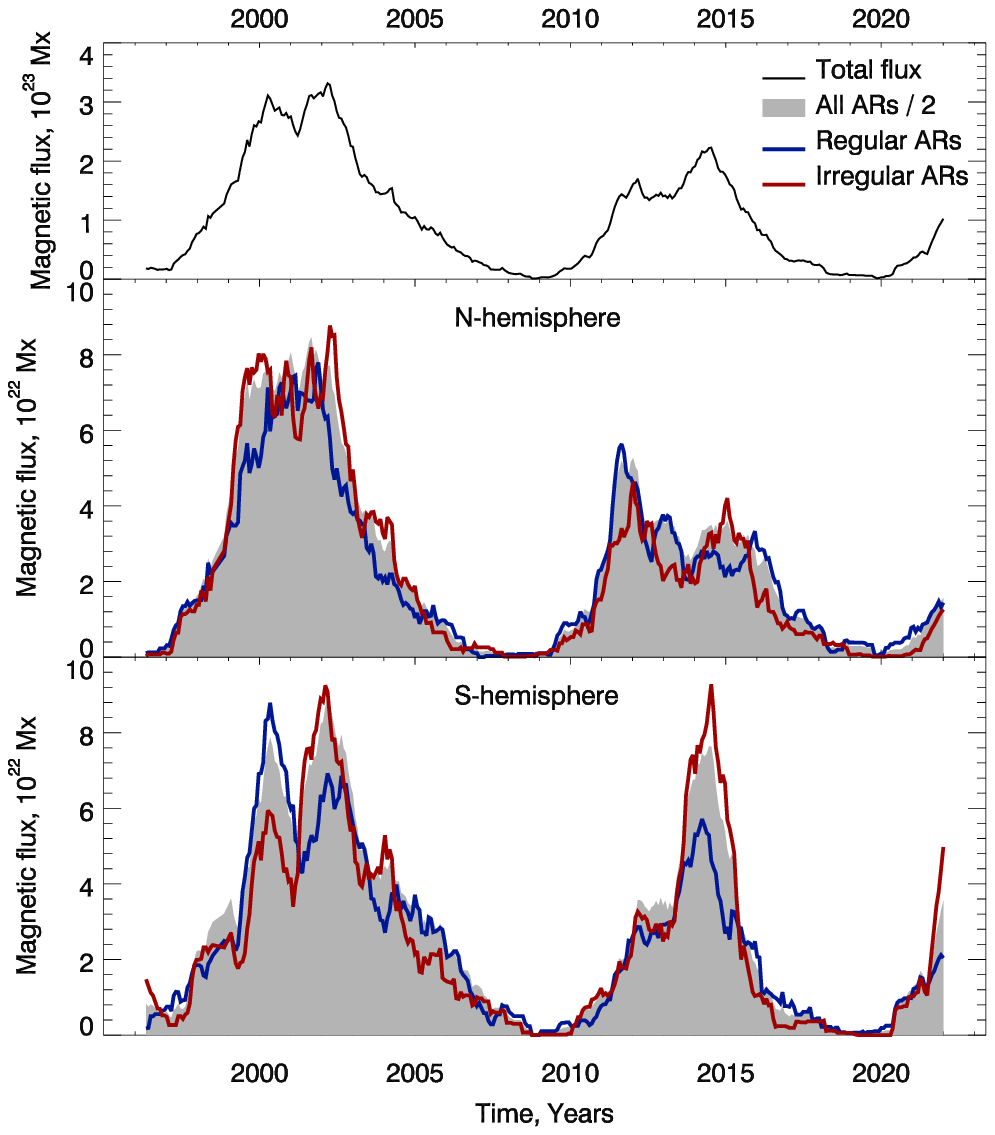}

\caption{Temporal variations of the fluxes of different-type ARs: total ARs (top panel, black line); 
regular ARs (blue line); irregular ARs (red line). 
The data for the N-(S-) hemisphere are presented in the middle (bottom) panel. 
The total flux distributed between the hemispheres is shown with gray fill.
All data are smoothed by thirteen-rotation moving average.}
\label{fig:temp}
\end{figure}

Temporal variations of the total unsigned magnetic flux (for all studied ARs) are presented 
in Fig.\,\ref{fig:temp} (top panel, black line).
The curve for the total flux illustrates the cycle progress and has the noticed double-peak structure, 
which is known since \citet{Gnevyshev63}.
In hemispheric data for the total flux (middle and bottom panels, grey fill), in the SC 23, 
the double-peak structure can also be traced.
However, the depth of the gap between two main maxima of the cycle varies in different hemispheres
(in the N-hemisphere, the time profile looks almost like a plateau, whereas, in the S-hemisphere, the flux 
decreases significantly).
In the SC 24, the pattern is different. 
The two main maxima are formed by ARs in different hemispheres (the N-hemisphere dominate in 
the first maximum, while the south fluxes more pronounced in the second maximum).
It is consistent with findings by \citet{Mandal16}, who showed that the double peaks may occur in one 
of the hemispheres without having any counterpart of the same in the other hemisphere.

Temporal profiles for regular and irregular ARs (Fig. \ref{fig:temp}, middle and bottom panels) are shown 
with blue and red lines, respectively. 
In each of the hemispheres, each of the profiles demonstrates rather a multi-peak structure.
Some of the peaks coincide with certain main maximum (maxima) of the cycle.
In different hemispheres, the peaks occur sometimes in-phase, and sometimes -- out-of-phase.

For different MMC-type ARs, the temporal profiles differ (Fig. \ref{fig:temp}). 
The regular ARs dominance is observed in the first maximum of the SCs 23 (S-hemisphere)
and SC 24 (N-hemisphere).
The irregular groups profiles demonstrate more peculiarities in the second maxima. 
Although the number of the irregular ARs is half the number of regular groups \citep{Abramenko18, Kashapova21},
their fluxes exceed the regular ARs fluxes in this temporal interval \citep{Abramenko23}. As it follows from Fig. \ref{fig:temp}, the dominance of the irregular ARs fluxes in the second maximum is provided by groups in the S-hemisphere.

Thus, the double-peak structure of a cycle as a whole is formed by ARs of both MMC types in both hemispheres.
The trends for the fluxes are more pronounced than that for the number of ARs  \citep[that was studied in our
previous research,][]{Zhukova23}.
The second main maximum of the cycle occurs due to ARs with irregularities in their magnetic configuration,
and the increased fluxes of the irregular ARs during this temporal interval are thought to be influenced 
by the turbulent component of the solar dynamo \citep{Abramenko23}.

The fact that the regular and irregular ARs fluxes are comparable to each other may also be interesting in terms of the surface polar field evolution, which eventually impact the predictability of the next solar cycle \citep{Jiang07, Karak12, Nagy17, Nagy19, Kumar22}. Specific polar surges are supposed to be formed due to the presence of large anti-Hale and non-Joy ARs at the solar surface, and this is confirmed by observations \citep[see, e.g.,][]{Mordvinov22}. Since the significant contribution of the irregular ARs in the solar cycle progress is found here, a comprehensive analysis of the relationship between characteristics of polar surges and the irregular ARs presence may be interesting as a topic for future studies. 

\section{Temporal variation of the asymmetry indices for the regular and irregular ARs}
\label{sec:ind}

To quantify the discrepancy in activity between the N- and S-hemispheres, we used several
asymmetry indices.
Along with the traditional normalized asymmetry index, $(N - S)/(N + S)$ \citep{Ballester05}, we also used 
the absolute asymmetry index, $(N - S)$, which is supposed to reproduce variations in activity 
better than the normalized index \citep{Temmer06}.
The point is that the absolute index shows the real imbalance between the hemispheres, whereas
the normalized index `spreads' it across the overall activity.
Usually, the relative asymmetry index demonstrates pronounced values in the minimum,
while the absolute asymmetry is strong in maximum  of the cycle \citep{Hathaway15}.
Thus, these two indices complement each other and we used both of them.
Note that when the index value is greater than zero, solar activity is dominant in the N-hemisphere, 
otherwise, the opposite is true.

As an additional way to calculate the hemispheric imbalance, we used the normalized index, 
$(N - S)^2/(N + S)$.
The choice was determined by the fact that this expression coincides with the equation for chi-square 
statistics for sunspot hemispheric data \citep[][see their Subsection 2.3]{Carbonell07}.
Strictly speaking, for the magnetic fluxes, the result of calculations by this formula cannot 
be accepted as a $\chi$-square statistics (this data are non-integer, and the sets of ARs are
too small in most of rotations).
However, some faint hint of the significance of the N-S asymmetry can be obtained using this index.

\begin{figure}

\includegraphics[width=\columnwidth]{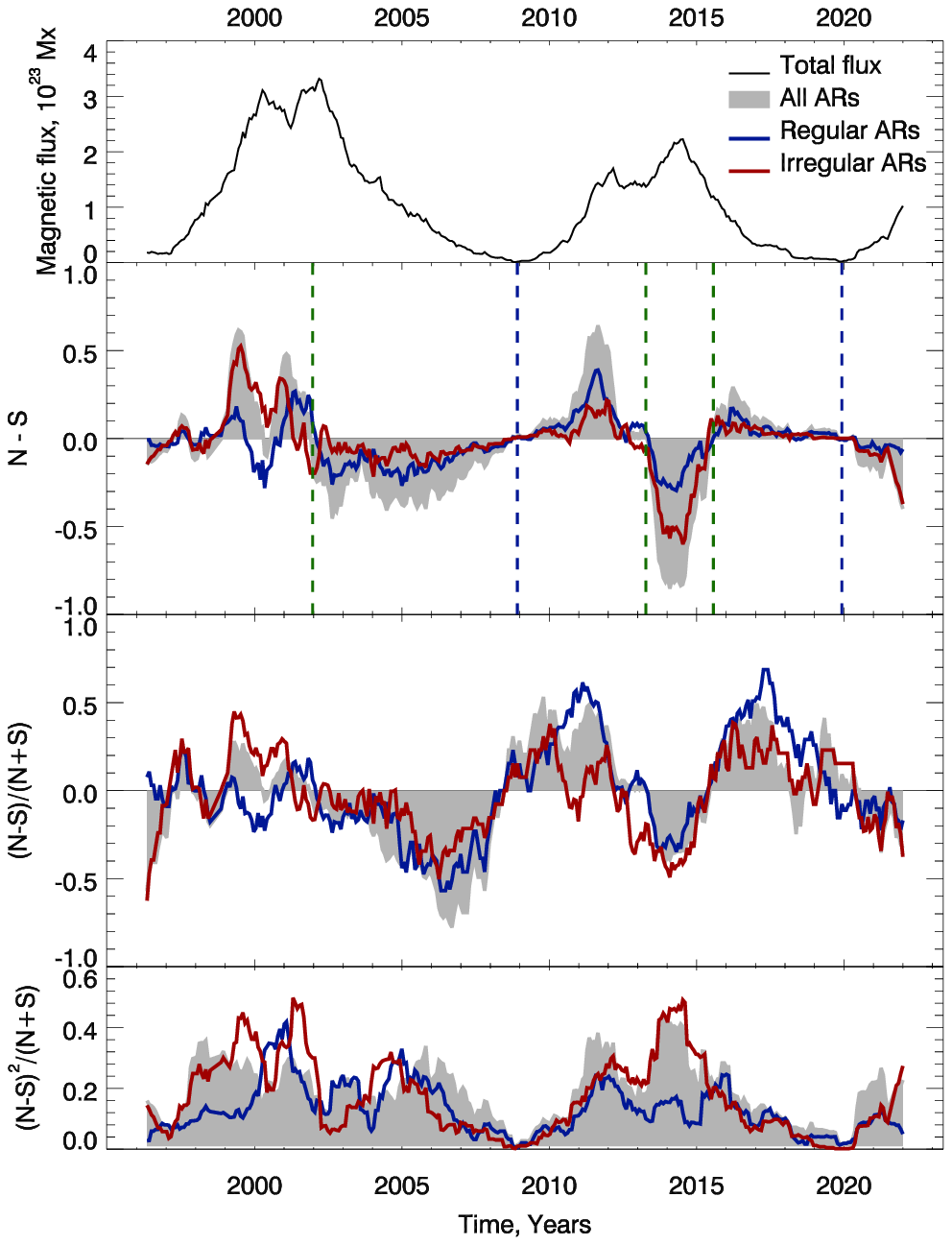}

\caption{Temporal variations of the total flux (top panel) and asymmetry indices for the flux data: 
absolute index, $N - S$, (second panel);  
traditional normalized index, $(N - S)/(N + S)$, (third panel);
normalized index $(N - S)^2/(N + S)$ (bottom) panel. 
All data are smoothed by thirteen-rotation moving average.
Dashed vertical lines (second panel) show the moments when the absolute asymmetry index sign changes in minima (blue) and maxima (green) of the cycles.
Notations are the same as in Fig. \ref{fig:temp}.}
\label{fig:ind}
\end{figure}

Temporal variations of the different asymmetry indices for the magnetic fluxes of total ARs (grey fill),
the regular (blue line) and irregular (red line) ARs are presented in Fig. \ref{fig:ind}.
The top panel is the same as in Fig. \ref{fig:temp} and provided to illustrate the cycle development.
The absolute asymmetry index, $N - S$, normalized indices, $(N - S)/(N + S)$ and 
$(N - S)^2/(N + S)$, are shown in the second, third and bottom panels, respectively. 

For total ARs, in the ascending phase and first maximum of the SC 23, one can see multiple transitions 
of activity from hemisphere to hemisphere.
During other time intervals, the asymmetry index retains its sign for several years.
In the SC 24, the sign change is observed twice, which is consistent with other studies 
\citep[see, e.g.,][]{Chowdhury19}.
The main points of the absolute asymmetry index sign changing (Fig. \ref{fig:temp}, second panel) 
are marked by the dashed vertical lines of different colors (blue for minima and green for maxima).

For the regular and irregular ARs, in the beginning of the SC23, the time profiles show considerable 
difference.
After 2002, the activity of both MMC type ARs follows the overall progress.
It is worth mentioning that, in some time intervals, the hemispheric asymmetry of the irregular ARs 
even more pronounced than that for the regular groups.
The absolute asymmetry index for the irregular ARs reaches its maximum values in intervals 
that can be called `extreme' for convenience (the SC 23, first maximum, N-hemisphere and SC 24, 
second maximum, S-hemisphere) (Fig. \ref{fig:ind}, second panel).
The values of the index in these intervals are greater than the highest value
for the regular groups (that is observed in the SC 24 in the first maximum).
Nevertheless, the values of the normalized index, $(N - S)/(N + S)$, are comparable for ARs of 
different magnetic morphology in the extreme intervals (third panel).
This may be due to the fact that these intervals fall at the maximum, where (as it was mentioned above)
the normalized index $(N - S)/(N + S)$ is not very pronounced.
In addition, the index $(N - S)^2/(N + S)$ allows us to assume the relevance of the strong 
hemispheric imbalance for the irregular ARs in the extreme intervals (bottom panel).
In the minima of the cycles, as the variations of the normalized index $(N - S)/(N + S)$ shows,  
a more noticeable N-S asymmetry is observed for the regular groups.

\section{Temporal variations of the `semi-sum' and `semi-difference' parts of the flux}
\label{sec:evenodd}

We also considered additional quantities that were conventionally called by us `semi-sum', $F_{SS}$, and `semi-difference', $F_{SD}$, parts of the magnetic flux. These quantities were obtained from the hemispheric data (using even and odd functions) as

\begin{eqnarray}\label{ss}
F_{\textbf{SS}} = \frac{(F_N + F_S)}{2},
\end{eqnarray}

\begin{eqnarray}\label{sd}
F_{\textbf{SD}} = \frac{(F_N - F_S)}{2},
\end{eqnarray}

where $F_N$ and $F_S$ are the parts of the flux observed in the N- and S-hemispheres, respectively. In a simple approximation (as discussed below in Section \ref{sec:dipquadr}), these quantities might be associated with the dipolar-like and quadrupolar-like components of the flux.

Since we operate with an unsigned magnetic flux (see Section \ref{sec:data}),
$F_N$ and $F_S$ must be assigned a certain sign. In accordance with the Hale polarity law and approach by \citet{Kitchatinov22}, in the SC 24, 
the flux was accepted as a positive (negative) in the N-hemisphere (S-hemisphere).
The sign was reversed in the adjacent minima.
For the minimum between SCs 23 and 24, 2008.03 (in the N-hemisphere) and 2009.02 (in the S-hemisphere) 
were adopted as the dates of sign reversal \citep[see,][their Table 1]{Kitchatinov22}.
For the minimum between SCs 23 and 24, the dates 2020.02 (in the N-hemisphere) and 2019.04 
(in the S-hemisphere) were fitted by the MMC ARs CrAO catalog data.

\begin{figure}

\includegraphics[width=\columnwidth]{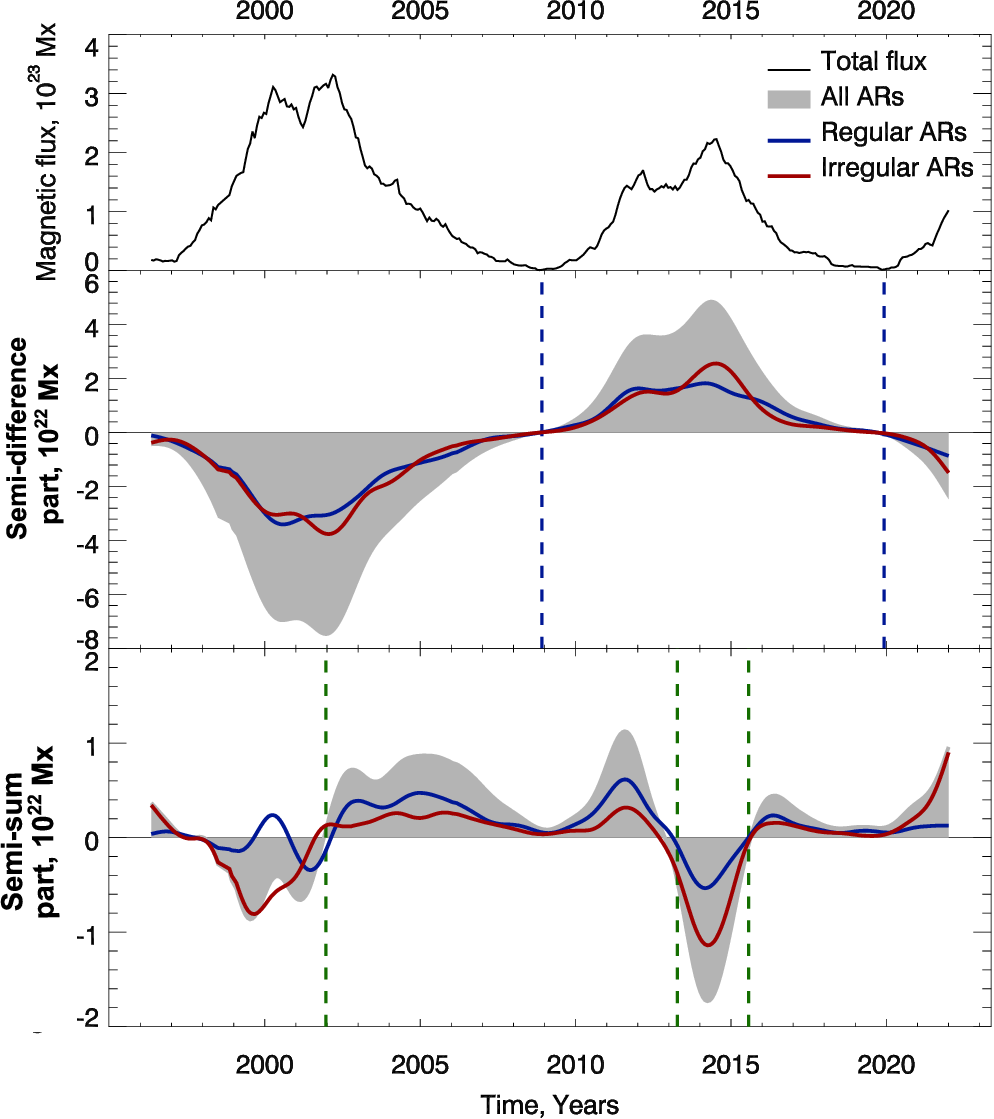} 

\caption{Temporal variations of the magnetic flux: total (top panel), semi-difference (middle panel) and semi-sum
(bottom panel) parts. The scales for different parts are not the same.
The data are smoothed using a Gaussian kernel (FWNM = 1.22 years).
Dashed vertical lines (for the moments of changing the sign of the absolute asymmetry index) are adopted from Fig. \ref{fig:ind}.
Notations are the same as in Fig. \ref{fig:temp}.
}
\label{fig:dipquadr}
\end{figure}

Temporal variations of the semi-difference (middle panel) and semi-sum (bottom panel) parts of the flux are shown in Fig. \ref{fig:dipquadr}.
The top panel is the same as in all previous Figs.
The data for the total ARs are represented by grey fill, time profiles for the regular and irregular ARs 
are shown by blue and red lines, respectively. To more accurately distinguish the peaks, these data are smoothed by a Gaussian kernel.
The full width at half maximum (FWHM) value for the kernel is accepted for 1.2 years. 
Dashed vertical lines (associated with the moments of changing the sign of the absolute asymmetry index) are adopted from Fig. \ref{fig:ind}.
These blue (for cycle minima) and green (for cycle maxima) lines almost coincide with the moments of changing the sign of the semi-difference and semi-sum parts of the flux, respectively.

For total ARs (Fig. \ref{fig:dipquadr}, grey fill), the temporal profile for the semi-difference part of the flux evidently reflects the course of the cycle (taking into account the sign changes). The semi-sum part is significantly inferior in strength to the dominant semi-difference part (note that the scales in the middle and bottom panels are not the same). The pattern for the semi-sum part is less regular and does not show typical cyclic variations. In 1999-2001, the regular and irregular ARs profiles varies in the out-of-phase manner.

For the regular and irregular ARs, the semi-difference part time profiles are pretty close  
(Fig. \ref{fig:dipquadr}, middle panel), however, the semi-sum part profiles (bottom panel) differ considerably.
For the semi-difference part of the flux, the regular ARs activity is somewhat more noticeable 
in the first maximum, whereas the irregular groups manifestation occurs in the second maximum.
And this is true for both cycles.
For the semi-sum part of the flux, the irregular ARs activity is most pronounced in the SC 23 (first maximum) and SC 24 (second maximum).
Wakening of the irregular ARs can also be guessed at the beginning of the SC 25.
Nevertheless, irregular ARs do not play a substantial role during the declining phase of SC 23 and the ascending phase of SC 24.
In general, irregular ARs show more dramatic changes in the semi-sum part of the flux than the regular groups. Possible interpretation is discussed below (see Sections \ref{sec:dipquadr}, \ref{sec:inter}).

\section{Correlation functions for the regular and irregular ARs}

\subsection{Auto-correlation for the semi-sum and semi-difference parts of the flux}
\label{sec:corrdq}

Our next test was aimed at finding signs of oscillations of the semi-sum and semi-difference parts of the flux produced by different MMC-type ARs.

The evidence of quasi-periodic variations in a time dependent variable $F(t)$ can be obtained using 
standart auto-correlation function

\begin{eqnarray}
C_F(n\delta t) = C_F(-n\delta t) = 
\frac{\displaystyle\sum_{i=1}^{N-n}{(F(t_i) - \overline{F})}{(F(t_{i+n}) - \overline{F})}}
{\displaystyle\sum_{i=1}^{N}{{(F(t_i) - \overline{F})}^2}} ,
\end{eqnarray}

where $\delta t$ is the time increment (one rotation in our case), $N$ is the total number of data points 
(341 rotations for the period of the study), $n$ is the time lag.

\begin{figure}

\includegraphics[width=\columnwidth]{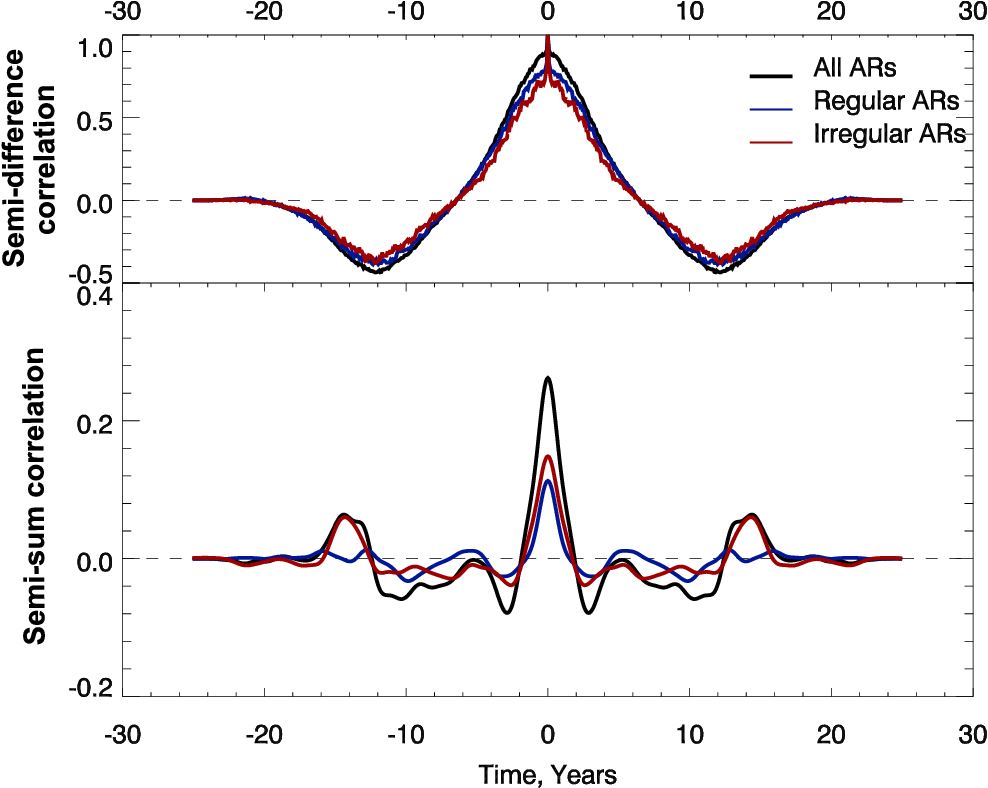} 

\caption{Autocorrelation function (Eq. 3) for the semi-difference (top panel) and semi-sum (bottom panel) 
parts of the flux.
The scales for different components are not the same.
The data are smoothed using a Gaussian kernel (FWHM = 1.22 years).
Notations are the same as in Fig. \ref{fig:temp}.
}
\label{fig:corrdq}
\end{figure}

The results of calculations according to Eq. 3 for the regular (blue lines) and irregular (red lines) 
ARs are shown in Fig. \ref{fig:corrdq}.
For the total flux, for the dominant semi-difference part (top panel, black line), the correlation function  
show a presence of cyclic variations with a period of 12 years. This is slightly different from the expected well-known value of 11 years for the solar cycle \citep{Hathaway15} and may be explained by the prolonged declining phase of the SC 24. For groups of different MMC types, the period of 12 years is also clearly pronounced.

For the semi-sum part of the flux (Fig. \ref{fig:corrdq}, bottom panel), the the correlation function is found to be more variable. For total ARs (black line), a period between 14 and 15 years can be detected. For the irregular ARs, the semi-sum part of the flux also demonstrates the close period and this is especially interesting, since for the regular ARs, a pronounced periodicity in the semi-sum part variations is not revealed.

\subsection{Cross-correlation between the hemispheric fluxes}
\label{sec:corrns}

We also found the correlation between the fluxes of ARs in different hemispheres as

\begin{eqnarray}
C_{NS}(n\delta t) = 
\begin{cases}
{\frac{1}{N-n}} \displaystyle\sum_{i=1}^{N-|n|} {F_N(t_{i+|n|})F_S(t_i),\,}& n < 0 \\
{\frac{1}{N-n}} \displaystyle\sum_{i=1}^{N-n} {F_N(t_i)F_S(t_{i+n})},\,& n \geq 0
\end{cases}
,
\end{eqnarray}
where time dependent variables $F_N(t)$ and $F_S(t)$ represent ARs in the N-hemisphere and S-hemispheres,
respectively.
Other notations are the same as for Eq. 3.
Note that in the Eq. 4, we deal with the convolution of fluxes in different hemispheres. 
Thus, we used non-normalized dimensional data.

\begin{figure}
\includegraphics[width=\columnwidth]{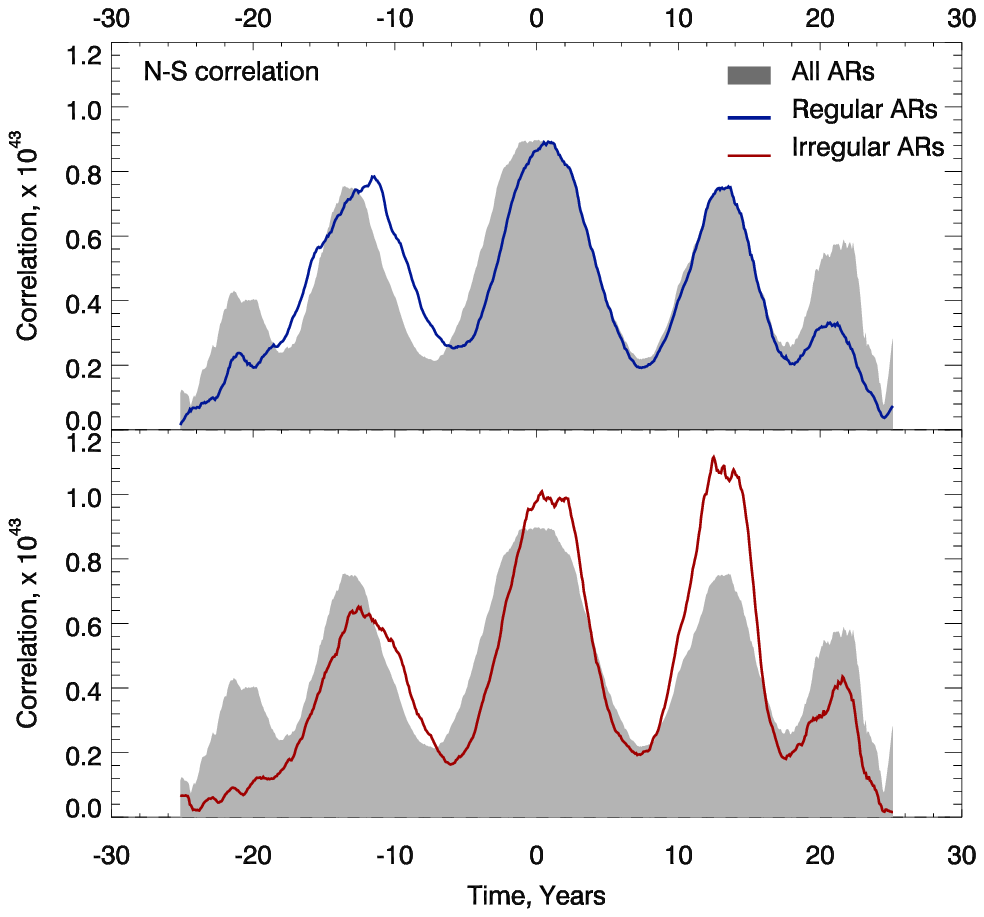} 

\caption{Correlation function (Eq. 4) for ARs fluxes in different hemispheres.
The data are smoothed by thirteen-rotation moving average.
Notations are the same as in Fig. \ref{fig:temp}.
}
\label{fig:corrns}
\end{figure}

In Fig. \ref{fig:corrns}, the results of calculations according to Eq. 4
for the regular (top panel, blue line) and irregular (bottom panel, red line) ARs are presented.
The right part of each of the graphs represents the case when the second series $F_S(t)$ is delayed 
relative to the first series $F_N(t)$.
In the left part, the time lag between the series is opposite. Thus, in each of the graphs, in the right part, the main side peak shows the relationship between the activity in the SC 23 (N-hemisphere) and SC 24 (S-hemisphere). The side peaks in the left parts correlate fluxes of ARs in the SC 23 (S-hemisphere) and SC 24 (N-hemisphere).

For total ARs (Fig. \ref{fig:corrns}, grey fill), a regular correlation pattern is observed.
The two main side peaks are slightly lower than the central one and indicate oscillations with a period 
between 11 and 12 years.
The left and right sides of the graph are almost symmetrical.

For the regular ARs (Fig. \ref{fig:corrns}, top panel), in the right part, the profile is approximately 
the same as for the total groups.
The blurring of the left side peak might be due to the mismatch of the time profiles for total and
regular ARs in the SC 24 (N-hemisphere) and in the declining phase of the SC 23 (S-hemisphere) 
(Fig. \ref{fig:temp}).

For the irregular ARs (Fig. \ref{fig:corrns}, bottom panel), on the left side of the panel,
the blurring is less pronounced.
However, in both parts of the graph, there is a significant difference in the height of the central and 
side peaks.
This represents a meaningful contrast between the graph and the usual pattern.
Note that the phenomenon is expressed specifically for the irregular groups.
A high right side peak implies a significant correspondence between the N-hemisphere (SC 23) and 
S-hemisphere (SC 24).
A small left side peak, on contrary, shows a weakened correlation between the N-hemisphere (SC 24) and S-hemisphere (SC 23).
Thus, for the irregular ARs, we deal with two different cases of the strengthened/weakened 
relationship between the two hemispheres in the adjacent cycles.

According to the theoretical concepts, a special symmetry between the adjacent cycles may be realized as the result of the   mixed-parity solutions for the dynamo models 
\citep{Sokoloff94}. The magnetic flux for the $n$-th SC in the N-hemisphere is predicted to be similar to the flux for the $(n+1)$-st SC in S-hemisphere and vice versa. As an example of such an interaction, the observed strong correlation between the hemispheres (N23--S24) could be considered.
Although, it should be noted that the case of weakening of correlation between the hemispheres (S23--N24) is also observed. The interaction of the dipolar and quadrupolar components of the global magnetic field was recently assumed as the reason for the N-S asymmetry expressed in features of the regular and irregular ARs statistics \citep{Zhukova23}. We consider the possibility to clarify this issue in the next Section \ref{sec:rel}.  

\section{The relationship between the dipolar and quadrupolar components of the large-scale magnetic field and occurrence of the regular and irregular ARs}
\label{sec:rel}

\subsection{Separation of the the dipolar-like and quadrupolar-like components from available flux data for the regular and irregular ARs}
\label{sec:dipquadr}

As it follows from the previous Subsection \ref{sec:corrns}, in terms of distinguishing between the regular and irregular ARs, it is interesting to infer the dipolar and quadrupolar components of the flux. In the current section, we consider such a possibility.

It is known that the large-scale magnetic field data are relevant for the harmonic analysis \citep[see, e.g.,][]{DeRosa12, Obridko21, Obridko23, Chu23}. Synoptic maps of the magnetic field are widely used for this purpose. Although the difference in the measurements of different instruments can lead to discrepancies that affect the energy recovered in each spherical harmonic mode \citep[see, e.g.,] []{Virtanen17, Finley23}, the advantage of this data type is sufficient number of harmonics, which is limited only by the map resolution. However, the synoptic maps do not allow us to identify the parts of magnetic flux originating from the regular and irregular ARs. Unfortunately, the sunspot data (suitable for this purpose) does not give the entire spatial pattern needed for the harmonic analysis. The synoptic-style mapping on the base of sunspot data \citep{Juckett03, Juckett06} has limitations and requires additional justification for the regular and irregular ARs. The currently available data for different MMC-type ARs imply a presence of a single signal from each of the hemispheres, which leads to the use of simplified expressions. 

An example of simplification is presented in \citet{Hazra19}, in their theoretical study of the relationship of the hemispheric asymmetry and parity flips in the dynamo mechanism,  where the global solar magnetic field was considered as the axisymmetric structure without omitting the non-axisymmetric components (see details in Appendix \ref{app}). We will refer to this assumption as the simple axisymmetric approximation further in the text. The similar method was applied by \citet{Munoz-Jaramillo13}, who used the dipolar and quadrupolar moments to improve solar cycle predictions based on the polar magnetic fields. 

In this approach, only axisymmetric terms ($m=0$) are taken into account (see Eq. \ref{dq}). This implies that the axial dipole and quadrupole assumed to be the main contributors to the magnetic field. Meanwhile, observational studies convincingly indicate the presence of significant non-axisymmetric components in the solar magnetic field \citep[see, e.g.,][their Fig. 1]{Hazra21}; the surface magnetic field appears to be complex and multipolar in the cycle maxima. When simplified approach is used, the influence of  non-axial and other higher harmonics  seemed to be blurred between the dipolar (Eq. \ref{DM}) and quadrupolar (Eq. \ref{QM}) parts without ability to distinguish the contribution of each term in Eq. \ref{dq}. Nevertheless, even some observational studies based on the synoptic maps data also did not divide the orders into axisymmetric zonal and non-axisymmetric modes and considered that all degrees $m$ of a given order $l$ are a whole entity \citep[see, e.g.,][]{Obridko21JASTP,Obridko23,Chu23}. Please note that, in terms of the dynamo theory, the separation of individual harmonics is a kind of abstraction. The dynamo process is characterized by significant non-linearity \citep[see, e.g.,][]{Hazra14, Passos14, Brun17, Charbonneau20, Charbonneau23, Karak23}. Dipolar and quadrupolar modes are known to be in continuous nonlinear interaction \citep{Hazra19}.  

In addition, observations can be interpreted in different ways. For instance, \citet{Kitchatinov22} considered the observed periods in the sunspot hemispheric asymmetry as the direct manifestation of the dipolar and quadrupolar modes. The expressions given by \citet[][ Eqs. 2]{Kitchatinov22} are similar to Eqs. \ref{DM}, \ref{QM},  which implies the axial symmetry of the dipolar and quadrupolar components and all related restrictions (discussed above). Another meaningful way is to present the hemispheric asymmetry as the result of superposition of the anti-symmetric and symmetric dynamo modes \citep{Schuessler18}. The assumption of this approach is the equality of the antisymmetric and symmetric modes, which, apparently, can be met only for critical dynamo modes \citep{Sokoloff94} and for some time intervals during the maxima \citep{Obridko21JASTP,Wang14}.

For the regular and irregular ARs, the approach by \citet{Schuessler18} cannot be applied because it requires more than a half of a century long observations of sunspot activity \citep[as in the work  of][which \citet{Schuessler18} relied on]{Ballester05}. We have in hands only two cycles long observations. The approach by \citet{Kitchatinov22} has a chance to be tried. The compiled semi-difference (Eq. \ref {sd}) and semi-sum (Eq. \ref{ss}) parts of the flux (Section \ref{sec:evenodd}) can be used as a proxy for the dipolar and quadrupolar components, respectively. However, doing that we should keep in mind all accompanying restrictions (see Appendix \ref{app} for further discussion).

The possibility of using such proxies for ARs of different magnetic morphology might be indirectly supported by the following signs. The temporal variations of the component of the dipolar parity (semi-difference part of the flux, Section \ref{sec:evenodd}, Fig \ref{fig:dipquadr}, middle panel) show the typical cycle progress for all studied ARs, as well as for the regular and irregular groups. The correlation analysis also shows the period similar to 11 years for all types of ARs (see Section \ref{sec:corrdq}, Fig. \ref{fig:corrdq}). Thus, the dipolar-parity part might be associated with the dipolar-like component of the flux.
Unfortunately, we cannot to make the similar firm inference about the quadrupole in the moment. 

\subsection{Interaction between the dipolar and quadrupolar components of the global magnetic field}
\label{sec:inter}

As a next step, we used the data on the asymmetry indices (Section \ref{sec:ind}) 
to develop ideas on 
the interaction between the dipolar and
quadrupolar components of the global magnetic field proposed in \citet{Zhukova23}.

In short, the global dynamo generates both the regular and irregular ARs \citep{Zhukova22MNRAS, Abramenko23}.
Irregularities occur due to the distortion of the magnetic flux tubes of ARs during their ascent 
through the convection zone \citep{Toriumi19}.
A turbulent component of the dynamo may be suggested as the reason for such distortions \citep{Abramenko21, Abramenko23}.
Its manifestations, expressed in an increase in the fraction of the irregular ARs, 
may be associated with the cycle development when the toroidal field (produced by the global dynamo) 
weakens and/or loses its regularity.
An increase in the number and fluxes of the irregular ARs in the second maximum of the cycle was found
in our previous research \citep{Abramenko18, Zhukova22GA, Abramenko23}.
A similar phenomenon is observed for the magnetic fluxes of the irregular ARs in this study 
(especially in the S-hemisphere, Fig. \ref{fig:temp}, bottom panel).

The hemispheric imbalance for the irregular groups may be explained by the additional weakening 
of the toroidal field due to the interaction between the dipolar and quadrupolar components 
of the magnetic field. 
Simple sketches for the component relationship for even and odd SCs can be found in 
\citet[][see their Figs. 5 and 6]{Zhukova23}. 
The idea is that \citep[as it follows from the classical magnetic cycle models][]{Babcock61, 
Leighton64, Parker55}, the global dipole  changes its orientation one time per 11-year cycle, whereas, in each SC, the quadrupole
magnetic field lines can be oriented in two opposite directions (and switch according to the quadrupole oscillation).
As a result, in each SC, in a given hemisphere, the total magnetic field is being 
strengthened or weakened due to the superposition of the dipolar and quadrupolar components.
This also applies to the toroidal field, since the quadrupole magnetic field lines 
are frozen into the plasma and are stretched along the equator by means of the differential rotation, as well as the dipole field lines. Please note that, although the global dipole reversals occurs near the solar maximum, at the photospheric level, the change of the toroidal field (manifested in the sunspot data) is observed with a delay, in the oncoming cycle minimum.

\begin{figure}
	
\includegraphics[width=\columnwidth]{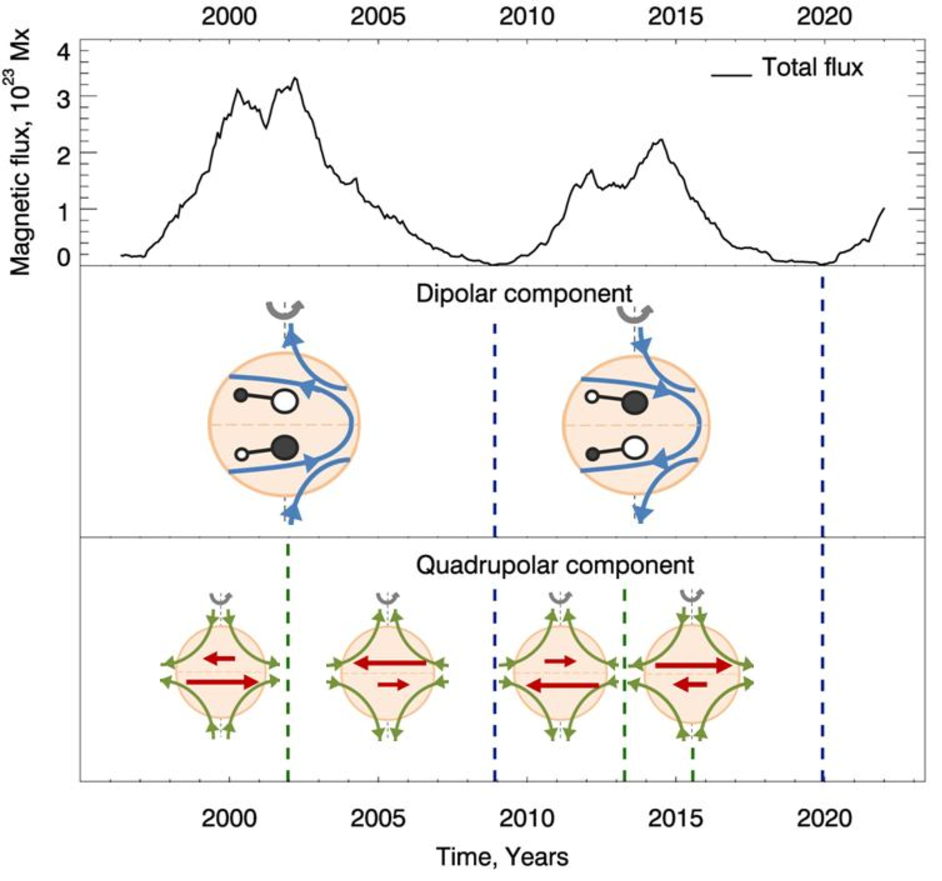}
	
\caption{Temporal variations of the total flux (top panel).
Sketches for dipolar (middle panel) and quadrupolar (bottom panel) components 
of the global magnetic field.
Magnetic field lines of the dipolar and quadrupolar components are shown as blue and green lines, respectively.
Red arrows represent the resulting field.
Regular ARs (for a given cycle and hemisphere) are shown with black (negative polarity spots) and white
(positive polarity spots).
Dashed vertical lines (for the moments of changing the sign of the absolute asymmetry index) are adopted 
from Fig. \ref{fig:ind}.}
\label{fig:sketch}
\end{figure}

In this study, we use the information both about the dipolar component orientation and 
the absolute asymmetry index sign to define a possible orientation of the quadrupolar component of the field.
In Fig. \ref{fig:sketch}, as in the previous figures, the top panel shows the cycle progress.
The middle panel demonstrate the orientation of the dipolar component of the field for a given cycle 
(blue lines).
The moments of the absolute asymmetry index sign reversal are adopted from Fig. \ref{fig:ind}
(dashed vertical lines).
Blue dashed lines correspond to the moments of changing the dipolar field orientation in the cycle minima.
Other dashed lines (representing other cases of the sign reversal corresponding to the cycle maxima) 
are seem to be associated with a change in the orientation of the quadrupolar component and are shown 
in green.

The bottom panel shows the results of our fitting of the quadrupolar component orientation (green lines).
The fitting was carried out as follows.
At first, we defined the total toroidal field (its strength and orientation).
The sign of the absolute asymmetry index (Fig. \ref{fig:ind}, second panel) indicates in which 
hemisphere the irregular ARs dominate in a given time interval.
The total field (Fig. \ref{fig:sketch}, bottom panel, red arrows) is supposed to be weakened 
in the corresponding hemisphere.
In addition, as the dipolar component exceeds the quadrupolar component in magnitude \citep[see, e.g.,]
[]{Sokoloff94}, the total field is assumed to be co-directed with the dipolar component.
The next step was to find out the direction of the quadrupole component of the field.
For this purpose, we considered both the direction of the dipolar component of the field and 
the strengthening/weakening of the total field.
In a given hemisphere, the dipolar and quadrupolar components are co-directional in the case 
of amplification of the total field, otherwise the opposite is true. 

According to this simple phenomenological model \citep[entirely corresponding to the classical ideas by]
[]{Babcock61, Leighton64, Parker55}, in the minima, when the direction of the dipolar component changes,
the quadrupolar orientation should remain unchanged. The moments of changing the quadrupolar component orientation are not determined by the 11-year cycle.

\subsection{Observational features for the dipolar-like and quadrupolar-like components of the flux}
\label{sec:feat}

To compare assumptions from the previous Subsection \ref{sec:inter} with the observation results, the interpretation from Subsection \ref{sec:dipquadr} was used. 

In the simple axisymmetric approximation, the semi-difference and semi-sum parts of the flux parts were considered as its dipolar-like and quadrupolar-like components, respectively. In terms of this approach, the fitted orientation of the quadrupolar component of the field in the sketch (Fig. \ref{fig:sketch}, bottom panel) appears to be in agreement with observed temporal variations of the  quadrupolar-like component of the flux (Fig. \ref{fig:dipquadr}, bottom panel). Changes in the direction of the magnetic field lines in the sketch occur almost simultaneously with changes in the sign in the graph.
Besides, the dipolar-like component reversals occur in the cycle minima, whereas the quadrupolar-like component direction varies at other times (near maxima), which is consistent with the model assumptions (Subsection \ref{sec:inter}). Please note that the quadrupolar component orientation (Fig. \ref{fig:sketch}) was fitted only for time intervals associated 
with the maxima of studied cycles.  

In addition, the dipolar-like and quadrupolar-like components representation used allow us to continue discussion of Figs. \ref{fig:dipquadr} and \ref{fig:corrdq} from a new perspective. As it follows from Fig. \ref{fig:dipquadr}, in the second maxima,
the irregular ARs show the enhancement
for both the dipolar-like and quadrupolar-like components of the flux (although in the SC 23, the quadrupolar peak is slightly shifted to the first maximum).
For the quadrupolar-like component, the peaks are more pronounced. It is also consistent with proposed simple phenomenological model. 

Our findings also meet indirect evidences in the other authors results obtained for the global magnetic field at the photospheric level and in corona \citep[where the masking effect of the higher harmonics on the dipolar and quadrupolar terms decreases, see][]{Wang14, Obridko24}.
For example, in 1999, the peak pronounced for both all and the irregular ARs (Fig. \ref{fig:dipquadr}) could appear due to increasing the amplitude of the quadrupolar-like component. Actually, at this time, the dipole and quadrupole strength becomes comparable  \citep[as it was found in][see their Fig. 2]{Wang14}. In 2012, we also can notice both the peaks in the quadrupolar-like part of the flux (Fig. \ref{fig:dipquadr}) and the immense growth in the non-axisymmetric quadrupole component of the global field according to the results by \citet{Wang14}. Apart of that, the time moments of the quadrupolar-like component peaking coincide with peaks in higher zonal harmonics $l=3,5$ \citep[see][]{Obridko21JASTP, Chu23}. It is also interesting that the irregular ARs increasing in maxima that was found for both the dipolar-like and quadrupolar-like components coincides with the moments of the equatorial (non-axisymmetric) global dipole manifestations \citep{Livshits06, DeRosa12, Obridko20MNRAS}. The rotation shearing may also converts the equatorial dipole into higher-order multipoles $l=3,5, 7,...$ \citep{Wang14}. The relationship between the local fields of sunspots and non-axisymmetric nature of the large-scale magnetic field might also be associated with the irregularities in ARs magnetic configuration. 

It should also be noted that the compiled quadrupolar-like component of the flux demonstrates the period of about 15-year (Fig. \ref{fig:corrdq}), which is close to the theoretical estimates \citep[][between
13 and 15 years]{Schuessler18}  and \citep[][16 years]{Kitchatinov21}. It is especially interesting that this periodicity is provided by the irregular ARs and is not detected for the regular groups. It might also indicate  a special role of the irregular ARs in the dynamo process.

\section{Concluding remarks}

The magnetic flux data for 3047 ARs from the MMC ARs CrAO catalog were used to study 
cyclic variations from May 1996 to December 2021 (complete SCs 23 and 24).
Along with the total flux, we analyzed the fluxes of subsets of ARs,
depending on their MMC type (the regular and irregular groups) and location relative 
to the equator (in the N- and S-hemispheres). We also considered the total flux of ARs as a proxy for
the subsurface toroidal flux and compiled its dipolar-like and quadrupolar-like components 
from ARs hemispheric data in the simple axisymmetric approximation.
The aim of this work was to study
the regular and irregular ARs hemispheric distribution and to clarify different MMC-type ARs involvement in the dynamo process. The features of the quadrupolar-like component of the flux expressed in the irregular ARs data were also considered.

As the result, we found the following.

(i) For ARs of each MMC type, in each of the hemispheres, time profiles demonstrate a multi-peak 
structure.
The most pronounced peaks are observed for the irregular ARs.
For groups of each type, in different hemispheres, the peaks occur sometimes in-phase and sometimes 
out-of-phase.
The double-peak structure  \citep[known since][]{Gnevyshev63} is formed 
by ARs of both MMC types in both hemispheres.

(ii) For both studied cycles, although the number of the irregular ARs is half less than the number 
of the regular groups \citep{Abramenko18, Kashapova21}, the irregular AR fluxes are comparable with ones of the regular 
groups (N-hemisphere) or exceed them (S-hemisphere, second main maximum).
As it was mentioned in the Introduction, the increase in the total irregular ARs fluxes in the second maximum is supposed to be due to the turbulent component of the dynamo \citep{Abramenko23}.
The pronounced flux hemispheric imbalance (found here) supports the previous hypothesis about weakening of the toroidal field and appearance of ARs-`violators' in one of the hemispheres 
due to the interaction between the dipolar and quadrupolar components of the magnetic field \citep{Zhukova23}.

(iii) Cyclic variations of asymmetry indices show that the N-S asymmetry of the irregular ARs is
even more pronounced than that of the regular groups.
In accordance with our simple phenomenological model, the absolute asymmetry index sign reversal occurs
when the mutual orientation of the dipolar and quadrupolar components of the magnetic field changes.
For the dominant dipolar component of the flux, as follows from the classical magnetic cycle 
models \citep{Babcock61, Leighton64, Parker55}, a sign reversal falls at minima.
In other cases, when the direction of the dipolar magnetic field lines remains unchanged, the asymmetry 
index sign reversal might occur due to the changes in the direction of the quadrupolar component.
Time profiles for the dipolar-like and quadrupolar-like components of the flux (inferred from observations in the simple axisymmetrical approximation) 
are consistent with these assumptions.

(iv) For the basic dipolar-like component of the total flux (as well as for different-type AR fluxes),
the auto-correlation function show a presence of cyclic variations with a period of 12 years.
A discrepancy with an expected value of 11 years \citep{Hathaway15} may be explained by the prolonged 
declining phase of the SC 24.
For the quadrupolar-like component, a period of about 15 years was found.
This is close to the theoretical estimates in \citet{Schuessler18} (between 13 and 15 years) and
in \citet{Kitchatinov21} (16 years).
Interestingly, that the period of 15 years is found only for the irregular ARs, 
whereas for the regular groups a pronounced periodicity is not revealed.

(v) For the total ARs and regular groups, for the adjacent cycles, a comparison of the hemispheric data   
using the cross-correlation function shows a standart correlation pattern with two side peaks (slightly inferior to the central peak in height).
For the irregular ARs the pattern is violated.
For the groups of this MMC type, the highest right side peak implies a strong correlation between the 
N-hemisphere (SC 23) and S-hemisphere (SC 24).
Since the mixed-parity solutions for dynamo models predict a special symmetry  \citep[similarity 
of activity in different hemispheres in cycles $n$ and $n + 1$, as discussed e.g. in][]
{Sokoloff94}, this is consistent with the theory.
However, the low left side peak shows a weak conformity between the S-hemisphere (SC 23) and 
N-hemisphere (SC 24).
Thus, for the adjacent cycles, during the transition of activity from the N- to S-hemisphere 
(and for the opposite case), we observe a strong (weak) correlation.

In summary, overall pattern for regular and irregular ARs in the cycle allows us to attribute their origin to the global dynamo action \citep{Abramenko18, Abramenko23}. A well-known phase lag between the hemispheric activity \citep[that finds an explanation in contemporary dynamo models, e.g.,][]{Norton14, Karak18ApJ, Hazra19, Kitchatinov22} was found for ARs of both MMC types. However, for the irregular ARs, the N-S asymmetry  show the features. Their increased activity, as well as considerable hemispheric imbalance, occurs in the second main maximum of the cycle, which is especially important.
The point is that the dynamo process involves two stages.
The first part of the cycle implies the transformation of the poloidal component of the global 
magnetic field to the toroidal one ($\Omega$-effect), and the differential rotation is
generally accepted as a trigger of this process.  
During the second part of the cycle, the poloidal component of the field restores from the toroidal one,
and the features of this process ($\alpha$-effect) are still being discussed. It might be assumed that the irregular ARs (pronounced in the second maximum) contribute in the $\alpha$-quenching, which could be different in different hemispheres. Our findings also confirm the crucial role of anti-Hale and non-Joy ARs for the polar field reversal that discussed, e.g., in \citet{Mordvinov22, Pal23}.
The irregular groups can modify the polar cap flux asymmetry and impact on the amplitude of the ongoing cycle, which is essential for the prediction models \citep{Jiang07, Karak12, Nagy17, Nagy19, Kumar22}. Thus, despite of the denomination (irregular, rogue, anomalous, etc.), ARs-`violators' are the integral part of the mechanism of solar activity and have special functionality in closing the dynamo loop. Apart of that, the irregular ARs are the source of strong flares and geoeffective events \citep{Abramenko21, Kashapova21}, which immensely impact the solar-terrestrial system \citep[see, e.g., the recent review by][and references therein]{Nandy23}. Please note that, on the one hand, the irregular ARs show significant fluxes comparable to those of the regular groups. On the other hand, present-day dynamo models are focused on the regular groups only. Thus, it would be interesting to consider the irregular ARs in the model design to complement our understanding of the solar cycle.
 
The irregular ARs also demonstrate a number of features for the  quadrupolar-like component of the flux (compiled in simplest axisymmetric approximation). The pronounced peaks in the maxima and  evidences of oscillations are found for them. The specific symmetry pattern (similarity of activity in different hemispheres in adjacent cycles) is also revealed for ARs of this MMC type. Although we could mention the possibility of a mix-parity dynamo solution \citep[see, e.g.,][]{Sokoloff94}, the simplicity of the approximation used restrict the degree of our certainty.
The same applies to  the possible relationship between the detected properties of ARs profiles and other observational features, such as evidences of the equatorial dipole \citep{Livshits06, DeRosa12, Obridko20MNRAS} and close amplitudes of the dipolar and quadrupolar components of the global field in certain temporal intervals in maxima \citep {Wang14, Obridko21JASTP}. Nevertheless, we emphasize that all the peculiar observational effects were found specifically for the irregular ARs.

In addition, implementation of the dipolar and quadrupolar modes with approximate equality of the components may leads to the global minima, although other reasons are also possible \citep[see, e.g., review by][]{Lopes14}. 
As \citet{Hazra19} showed, the solar cycle might have resided in quadrupolar parity states in the past, 
which provides a possible pathway for predicting parity flips in the future.
It is quite possible that, with the dominance of the quadrupolar component, all ARs will appear 
in only one hemisphere \citep[as in the vicinity of the Maunder Minimum,][]{Wideburg09, Hayakawa21}, 
and they will mostly be irregular.
We thus speculate that a significant increase in the fraction of the irregular ARs during future cycles
may warn us about critical changes in the level of solar activity.
 
Anyway, all aspects of interpretation are the subject of theoretical researches, which are beyond the scope of this article. We present here only observational results about the magnetic fluxes of ARs with different magnetic morphology. 

\section*{Acknowledgements}

The author is grateful to the anonymous reviewer for his/her meaningful comments that made it possible to improve the article. The author is thankful to V.I.Abramenko for her valuable remarks and to R.A.Suleymanova for the data on the SC 23 in the MMC ARs CrAO catalog. The author would also like to thank V.N.Obridko, D.D. Sokoloff, L.L.Kitchatinov, M.S.Butuzova, S.A.Korotin, K.N.Grankin, S.M.Andrievsky for fruitful discussions. The study was financially supported by the Russian Ministry of
Science and Higher Education, agreement \textnumero 122022400224-7. 

\section*{Data Availability}

The MMC AR CrAO catalog is available at the CrAO web site (https://sun.crao.ru/databases/catalog-mmc-ars).
Additional comments can be recieved from the author on request. 



\bibliographystyle{mnras}
\bibliography{Zhukova} 


\appendix

\section{Defining the dipolar and quadrupolar components of the magnetic field from observations}
\label{app}

In general, the radial magnetic field on the solar surface $B(\theta,\varphi)$ can be represented in terms of spherical harmonics $Y_l^m(\theta,\varphi)$ as

\begin{eqnarray}\label{field}
	B(\theta,\varphi, t) = 
	\displaystyle\sum_{l=0}^{\infty}{\displaystyle\sum_{m=-l}^{l}{B_l^m(t)Y_l^m(\theta,\varphi)}},
\end{eqnarray}

where $\theta$ and $\varphi$ are polar (co-latitudinal) and azimuthal (longitudinal) coordinates, respectively \citep[see, e.g.][]{Chu23}. Time-dependent complex coefficients $B_l^m(t)$ can be found from the orthogonality relationship by the integration over the surface of a sphere, for example, as

\begin{eqnarray}
	\int{Y_{l'}^{m'*}(\textbf{r})Y_l^m(\textbf{r})d\Omega = \delta_{ll'}\delta_{mm'}}.
\end{eqnarray}

The spherical harmonics $Y(\theta,\varphi)$, which represent the angular portion of the solution to Laplace's equation in spherical coordinates, can be expressed as

\begin{eqnarray}
	Y(\theta,\varphi) = C_l^mP_l^m(\cos\theta)e^{im\varphi},
\end{eqnarray} 

where $P_l^m(cos\theta)$ are associated Legendre polynomial of order $l$ and degree $m$. Coefficients $C_l^m$ are defined as

\begin{eqnarray}
	C_l^m = (-1)^m\Bigl[\frac{2l+1}{4\pi}\frac{(l-m)!}{(l+m)!}\Bigr]^{1/2}.
\end{eqnarray} 

In practice, the number of terms in Eq. \ref{field} is finite and depends on the type of the observational data. Synoptic maps are widely used as the basis for calculations, and truncation limit $l_{max}$ depends on maps characteristics \citep[see, e.g.,][]{DeRosa12, Obridko21, Finley23}. Note that  coefficients $B_l^m(t)$ have a complex nature, and the amplitudes of the spherical harmonics modes appear in the real part (for $m>0$) and imaginary part (for $m<0$) \citep{DeRosa12}. Thus, one can taking into account a symmetry between spherical harmonics with orders $m$ and $-m$ (for a given value of $l$) and start the sum at $m=0$ (instead of $m=-l$). Therefor, the Eq. \ref{field} takes a form 

\begin{eqnarray}\label{harms}
	B(\theta,\varphi, t) = 
	\displaystyle\sum_{l=0}^{l_{max}}{\displaystyle\sum_{m=0}^{l}{B_l^m(t)Y_l^m(\theta,\varphi)}}.
\end{eqnarray} 

Limiting only the dipolar and quadrupolar components of the field (lower-order terms, $l=1$ and $l=2$, respectively) and rewriting the expression in more detail (to examine its structure), one can obtain from Eq. \ref{harms} the following expression

\begin{eqnarray}\label{dq}
	B(\theta,\varphi,t) = \underbrace{\sqrt{\frac3{4\pi}}B_1^0(t)\cos\theta-\sqrt{\frac3{8\pi}}B_1^1(t)\sqrt{1-\cos^2\theta})e^{i\varphi}}_{\text{first-order terms}}+ \nonumber\\
	{\small\text{second-order terms}}
	\begin{cases}
	+\sqrt{\frac5{4\pi}}B_2^0(t)\cdot\frac12(3\cos^2\theta-1)- \\
	-\sqrt{\frac5{24\pi}}B_2^1(t)\sqrt{1-\cos^2\theta}\cdot\cos\theta\cdot e^{i\varphi}+\\
	+\sqrt{\frac5{96\pi}}B_2^2(t)\cdot3(1-\cos^2\theta)\cdot e^{2i\varphi}.
	\end{cases}
\end{eqnarray}

In Eq. \ref{dq}, each of the parts (for each of the orders) consists of both axisymmetric ($m=0$) and non-axisymmetric ($m>0$) terms. In the first-order part ($l=1$), the first term represents an axial global dipole, which axis approximately coincides with the rotation axis of the Sun, while the second term can be associated with a horizontal (equatorial) dipole \citep[][]{Livshits06, DeRosa12, Wang14,  Obridko20MNRAS}. The second-order part contains the quadrupolar ($l=2$) terms. The first (axisymmetric, $m=0$) term is responsible for a zonal harmonic, while the non-axisymmetric terms, $m=l=2$ and $m \neq l$, are related to the sectorial and tesseral structures, respectively \citep{Zieger19, Mikhaylutsa20, Obridko24}.

In addition, the terms in both-order parts (Eq. \ref{dq}) can be classified as antisymmetric (odd $l+m$) and symmetric (even $l+m$) with respect to the equator. According to different nomenclatures, these families of harmonic modes also referred as `primary-family' (`dipolar') and `secondary-family' (`quadrupolar'), respectively. It is important that the nomenclature dipolar/quadrupolar (used in the sense of equatorial symmetry) can lead to confusion, and the equatorial dipole ($l=1$, $m=1$) can formally be assigned to the `quadrupolar' family \citep[see][and references therein for more details]{DeRosa12}.

Please note that the axisymmetric terms ($m=0$) in the Eq. \ref{dq} do not contain a dependence on the azimuthal angle and have a simple form that facilitates further transformations. As it is shown in \citet{Hazra19}, in the case of axial symmetry (assuming the axial dipolar and quadrupolar moments the main determinants of the field), we can obtain for a particular latitude for the field in the N-hemisphere 

\begin{eqnarray}\label{N}
	B_N = C_1\cdot DM\cdot\cos\theta + C_2\cdot QM\cdot\frac12(3\cos^2\theta-1), 
\end{eqnarray}

and for the field in the S-hemisphere

\begin{eqnarray}\label{S}
	B_S = -C_1\cdot DM\cdot\cos\theta + C_2\cdot QM\cdot\frac12(3\cos^2\theta-1),
\end{eqnarray}

where coefficients $B_1^0$ and $B_2^0$ are expressed as the dipolar ($DM$) and quadrupolar ($QM$) moment, respectively. Coefficients $C_1^0$ and $C_2^0$ are also designed: $C_1\equiv C_1^0=\sqrt{\frac3{4\pi}}$;  $C_2\equiv C_2^0=\sqrt{\frac5{4\pi}}$. In addition, in Eqs. \ref{N}, \ref{S}, different signs are assigned to the antisymmetric component of the field in different hemispheres, whereas the symmetrical component has the same sign on different sides of the equator.

With respect to the dipolar and quadrupolar moments, the combination of Eqs. \ref{N} and \ref{S} is a simple linear system. It provides the following solutions  

\begin{eqnarray}\label{DM}
	DM = \frac1{2C_1\cos\theta}(B_N-B_S), 
\end{eqnarray}

and

\begin{eqnarray}\label{QM}	
		QM = \frac1{C_2(3\cos^2\theta-1)}(B_N+B_S).
\end{eqnarray}

Thus, in the axisymmetric approximation, the dipolar and quadrupolar components of the field turn out to be proportional, respectively, to the difference and sum of the hemispheric signals.

\citet{Hazra19} used Eqs. \ref{DM} and \ref{QM} to define a parity function in terms of dipolar and quadrupolar moments to reveal the relationship between the solar parity reversal and hemispheric asymmetry. The close approach in the calculation of the dipolar and quadrupolar moments was used by \citet{Munoz-Jaramillo13} when improving the solar cycle predictions based on the northern and southern polar magnetic field data.  A similar method proved to be useful when working with observational data limited to a single signal from each of the hemispheres at any given temporal interval \citep[see, e.g.,][who deals with sunspot area data in his study of the dipolar and quadrupolar dynamo modes]{Kitchatinov22}.


\bsp	
\label{lastpage}
\end{document}